\RequirePackage{fix-cm}
\documentclass{svjour3}                     % onecolumn (standard format)
\smartqed  % flush right qed marks, e.g. at end of proof
\usepackage{graphicx}
\begin{document}

\title{Effect of milk fat content on the viscoelasticity of mozzarella-type cheese curds%\thanks{Grants or other notes
}

%%\subtitle{Do you have a subtitle?\\ If so, write it here}


\author{Hiroyuki Shima \and Morimasa Tanimoto}

\institute{Hiroyuki Shima \at
           Department of Environmental Sciences, University of Yamanashi, 4-4-37, Takeda, Kofu, Yamanashi 400-8510, Japan \\
           Tel.: +81-55-2208834\\
           Fax: +81-55-2208834\\
           \email{hshima@yamanashi.ac.jp} %  \\
       \and
           Morimasa Tanimoto \at
           Department of Local Produce and Food Sciences, University of Yamanashi, 4-4-37, Takeda, Kofu, Yamanashi 400-8510, Japan
}

\date{Received: date / Accepted: date}

\maketitle

\begin{abstract}
The effect of fat content in cheese curds on their rheological properties
were examined using dynamic shear measurements.
Surplus fat addition to milk samples
caused two distinct types of changes in the temperature dependence
of the viscoelastic moduli of resultant curds.
The first was a significant reduction in the moduli over a wide temperature range,
which is attributed to the presence of liquefied fat globules within
the milk protein network.
The second was the excess contribution to the low-temperature moduli owing to the reinforcing effect of solidified fat globules.
An upward shift in the sol-gel phase transition temperature
driven by an increased fat content was also observed.
\end{abstract}

%************************
\section{Introduction}
\label{}

Fat and protein are the two primary components in raw milk.
The fat content of bovine milk is nearly 4\% by weight,
and it is dispersed in milk serum as globules with diameters
that range from 0.2 $\mu$m to 15 $\mu$m, c.a. 4 $\mu$m on average \cite{Michalski2004}.
Similar to fat globules, casein proteins (i.e., the major class of milk protein) exist
as colloidal particles, known as casein micelles,
and have diameters that range from 50 nm to 500 nm (average 120 nm).
These colloidal domains comprise almost 80 \% of the total solid content in milk.
Therefore, their structural stability and inter-particle interactions
strongly affect the quality of dairy products such as cheese, yoghurt, and butter.
In particular, the presence of fat in cheese is necessary to develop
the characteristic flavour profile and favoured mouth-feel.

The production of natural cheese is initiated
by the addition of rennet to milk.
The rennet-induced proteolysis of the surfaces of casein micelles
leads to their aggregation, resulting in a three-dimensional protein network.
The network exhibits non-uniform viscoelasticity in accordance with changes in
the temperature, pH, and protein concentration \cite{Nabulsi,Catarino}.
Cavities in the network are filled with fat globules and some whey;
the total mixture of these materials comprises a cheese curd.
Many fat globules in the curd remain stored, even after curd syneresis is completed,
and they contribute to the desirable functional properties of the final cheese product.
In fact, artificial removal of fat from the curd
causes quality degradation,
leading to a firm and dry cheese that melts poorly \cite{Mistry2001}.
Toward quality improvement, numerous studies have focused on
the effect of fat content or its reduction in cheese.
Despite consumer enthusiasm for fat-free diets,
these attempts have met with limited success \cite{Banks,Childs,Skeie}.
A better understanding of the interplay between fat globules and the protein network
is indispensable for developing a solution.

Aside from the practical motivation,
it is also interesting from an academic perspective
to explore the effects of fat content on the rheology of cheese curds.
An important feature of fat globules,
which contribute to curd rheology,
is the wide variety in in size and melting temperature.
The broad distribution of fat globule sizes allows them to
interact with cheese microstructures in multiple ways.
Large fat globules are likely to disrupt a portion of the protein network and
suppress direct cross-linking between protein threads.
Hence, if they are liquefied,
large globules are expected to plasticize adjacent protein threads \cite{Johnstona1984},
yielding a structurally loose matrix with reduced firmness.
In contrast, small globules tend to occlude the fine empty spaces in the network \cite{MichalskiLait2004}
and are thought to act as reinforcing fillers \cite{Desai1994} if they are in solid phase.
However, a simple explanation of the temperature dependence of the fat content
may be insufficient owing to the wide variety of fat melting points.
There is not a sharp difference between the liquid and solid states of fat globules in curds.
A single fat globule encloses many kinds of triglyceride isomers with different
melting points \cite{Jensen1991,Lopez2005},
and thus the solidity and fluidity of the globule are determined by the relative proportion of isomers.
The actual melting temperature ranges from $-40 ^\circ$C to 40$^\circ$C,
between which crystalline and liquid fat coexist in curd \cite{Pilhofer1994}.
It remains unclear how the two competing roles of fat globules,
as plasticizers and reinforcing fillers, are manifested
with respect to thermal-induced changes in cheese curd rheology.

%%%--------------------------------------------------------------
\begin{figure*}[ttt]
\centering
\includegraphics[width=0.3\textwidth]{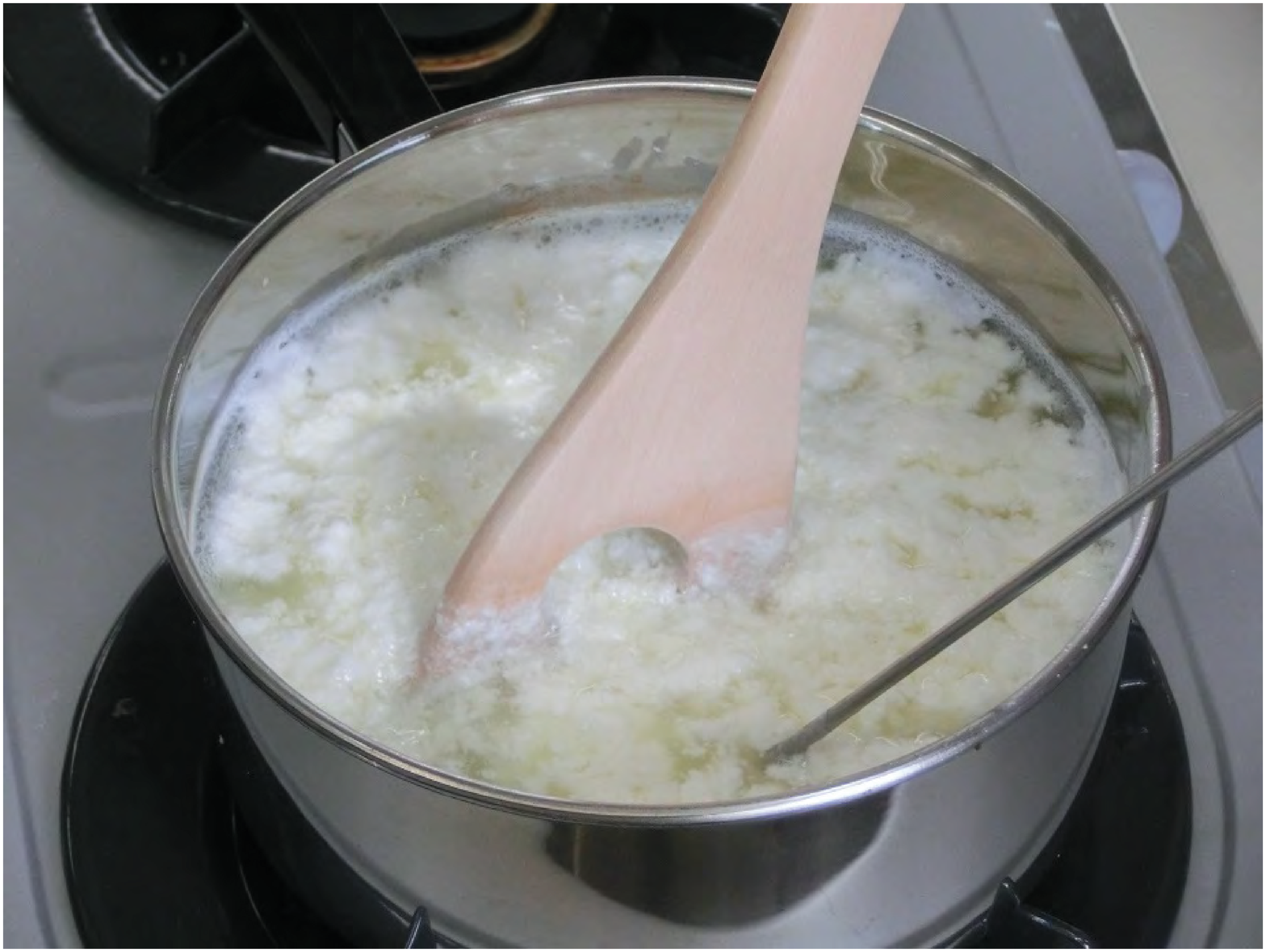}
\includegraphics[width=0.3\textwidth]{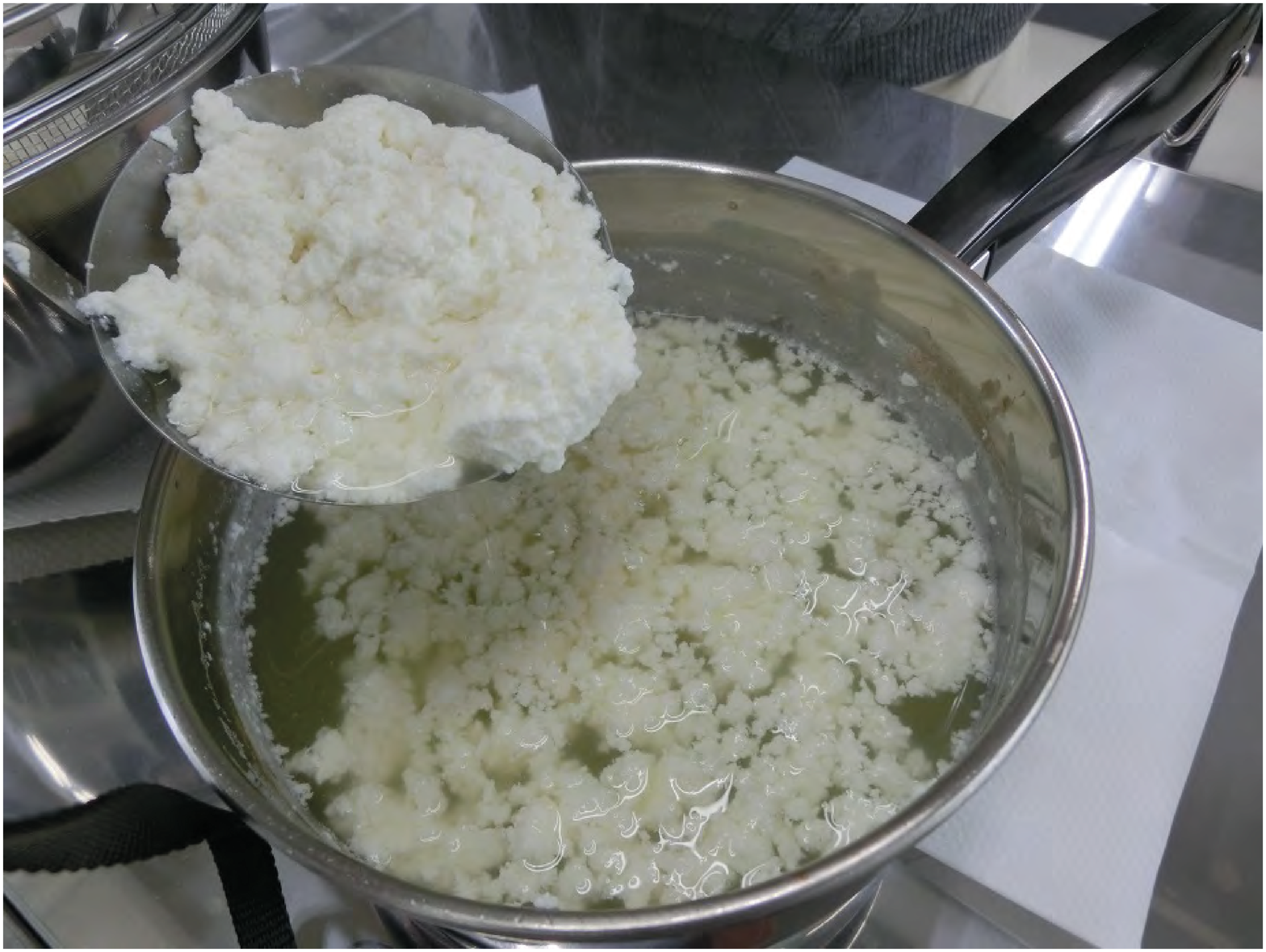}
\includegraphics[width=0.3\textwidth]{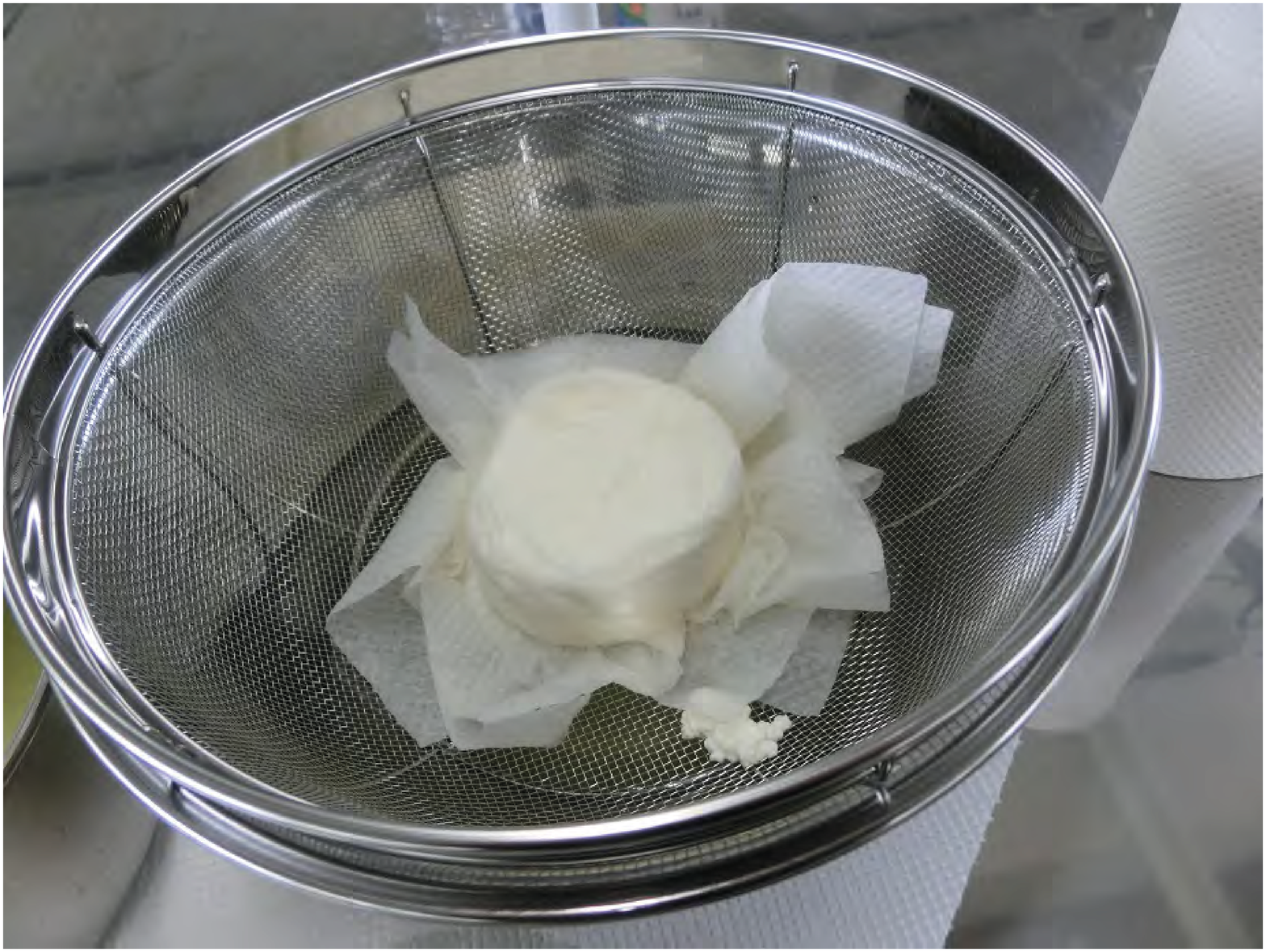}
\caption{Left: Initiation of milk clotting obtained just after adding rennet.
Middle: Curd granules. Right: Final state from which all whey is removed.}
\label{fig_milk}
\end{figure*}
%%%--------------------------------------------------------------

In the present study, we address the effect of fat content and pH control
on the viscoelastic moduli of rennet cheese curds.
The pH control allows us to examine the effects of fat content
under various structural conditions of the protein network.
High pH conditions cause protein networks to become weaker and more porous.
In contrast, low pH conditions result in network contraction,
in which either or both of the effects as plasticizers or fillers may be enhanced.
To verify our conjecture, we performed
dynamic shear tests and measured the variation in the temperature dependence of the moduli with changes in pH.
Particular emphasis was paid to the rheological behaviours below 20 $^\circ$C and above 50 $^\circ$C,
wherein most fat globules are solidified and liquefied, respectively.

\section{Material and method}

\subsection{Preparation of sample milk}

\begin{figure*}[ttt]
\centering
\includegraphics[width=0.7\textwidth]{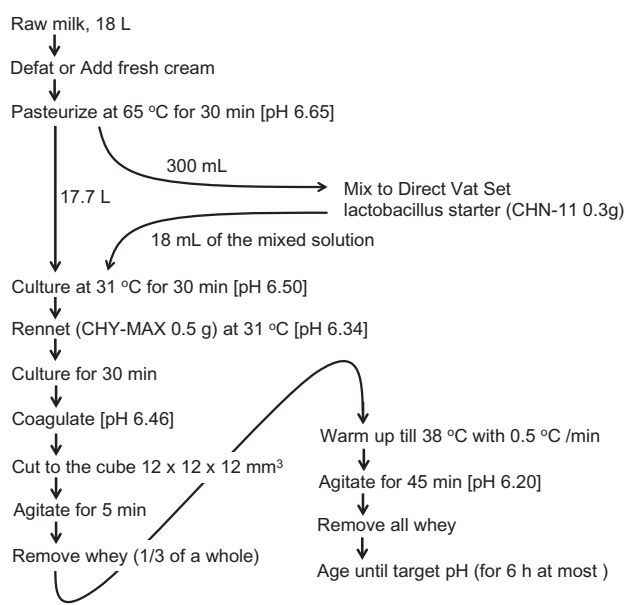}
\caption{Flow chart of mozzarella-type cheese curd production.}
\label{fig_diagram}
\end{figure*}

Figure \ref{fig_diagram} displays a flow chart summarizing the production of pH- and fat-controlled cheese curds.
Raw milk was obtained from the Kiyosato Milk Plant located at the foothills of Mount Yatsugatake, Japan.
To assess the effects of fat content on cheese rheology,
two classes of fat-adjusted milk were prepared.
Skim milk was produced with a fat content of less than 1\% by weight.
Fat-enriched milk was produced by adding 1 kg of fresh cream with 47 \% fat
into 17 kg of raw milk with 3.8 \% fat.
The fat-enriched milk contained 6.2 \% fat by weight.
Each milk sample (18 L) was first pasteurized by maintaining the sample at 65 $^\circ$C for 30 min.
This process eliminates bacteria,
thus preventing the degradation of milk proteins at high temperatures.
After pasteurization, the samples were then cooled to 31 $^\circ$C.
The sample pH was 6.65 at this stage, and was directly determined using
an electrode-type pH meter (SK-620PH, skSATO, Tokyo, Japan).

\subsection{Starter insertion}

Milk acidification was triggered by adding
18 mL of a Lactobacillus culture solution to the milk sample.
The culture solution was a mixture of
0.3 g of Direct Vat Set (DVS) Lactobacillus starter
(CHN-11, Chr. Hansen, Nosawa \& Co., Ltd., Tokyo, Japan)
with 300 mL of pasteurized milk that was prepared in advance.
After adding the culture solution, the sample was maintained at 31 $^\circ$C,
which is the temperature that produces the optimal Lactobacillus activity,
for 30 min. The sample pH at this stage was 6.50.

%%%--------------------------------------------------------------
%%%--------------------------------------------------------------

\subsection{Cheese curd production}

Rennet was then added to the above sample that was slightly acidified by lactobacilli.
In this experiment, 0.5 g of rennet (CHY-MAX, Chr. Hansen, Nosawa \& Co., Ltd.)
dissolved in sterile cold water was added to the sample,
which was then maintained at 31 $^\circ$C for 30 min.
After the milk started to coagulate, the sample was cut into cubes of
$12\times 12 \times 12$ mm$^3$ to remove a portion of whey from the curds.
After cutting, the sample was gently agitated for 5 min to encourage the removal of whey.
As a result, whey corresponding to one-third of the original sample weight was eliminated,
yielding two classes of curd granules that differed in fat content.

To complete the whey removal process, hot water was added gradually to the curd granules
so that the sample temperature increased at a rate of 0.5 $^\circ$C/min.
When it reached 38 $^\circ$C, the granules were gently agitated again for 45 min;
eventually, all whey was removed.
In the final step, the curds were aged until they reached the target pH (4.8-5.7).
The time required to attain curds with the lowest pH was approximately 6 h.
After pH adjustment, a series of curds with different pH values were frozen and stored.
Immediately before measurement, the curds were defrosted in a refrigerator
and then stirred at 50-60 $^\circ$C.

\subsection{Composition analysis}

The chemical composition of the fat-controlled curds is summarized in Table \ref{table01}.
For both no-fat and high-fat samples, the highest pH among the samples examined are displayed,
as well as the analysis method.
It is noteworthy that the amount of fat was supressed to 2.7 g/100 g in the no-fat sample,
while it increased to 30.1 g/100 g in the high-fat sample.
The ratios of calcium to protein are presented in the bottom row of Table \ref{table01},
and indicate that fat control has no effect on the calcium content within the protein network.

\begin{table*}[ttt]
\caption{Chemical compositions of the obtained curds with different fat contents.
($^*$ICP = Inductively Coupled Plasma)}
\centering
  \begin{tabular}{lrrc} \hline
       & No-fat  & Hi-fat & Method of analysis \\ \hline %\hline
    pH & 5.60 & 5.70 & pH meter \\
    Fat (g/100g) & {\bf 2.7} & {\bf 30.1} & Acid hydrolysis method \\
    Protein (g/100g) & 38.8 & 16.3 & Kjeldahl method \\ [8pt]
    Moisture (g/100g) & 53.1 & 48.2 & Atmospheric heating drying method \\
    Ash (g/100g) & 3.7 & 1.7 & Direct ashing method \\
%%    Carbohydrate (g/100g) & 1.7 & 3.7 & \\
%%    Energy (kcal/100g) & 186 & 351 & \\
    P (g/100g) & 0.805 & 0.346 & $^*$ICP atomic emission spectroscopy \\
    Ca (g/100g)& 1.23 & 0.502 & same as above \\
    Ratio Ca/Protein ($\times 10^{-2}$) & 3.17 & 3.08 &  \\ \hline
  \end{tabular}
\label{table01}\end{table*}

\subsection{Dynamic shear measurement}

Viscoelastic moduli of the cheese curds and their dependence on pH, temperature, and fat content
were evaluated by small-amplitude oscillatory shear tests.

These are non-destructive tests for determining the viscoelasticity 
of a material \cite{Gunasekaran,Tunick2011},
and have been widely used for analysing cheeses and other foodstuff
such as chocolates \cite{Vaart2013} and rice bran \cite{YHZhang2014}.

Specifically, an oscillatory shear strain is applied to the sample
at constant frequency of 1 Hz and a constant strain amount of 0.1 \%,
which satisfies the linear viscoelastic condition.
Decreasing the temperature from 65 $^\circ$C to 5 $^\circ$C
was carried out in a ramp fashion with a constant cooling rate of 2 $^\circ$C/min.
The observed quantities were the temperature ($T$) dependences of
the elastic (or storage) modulus, designated $G'(T)$
and the viscous (or loss) modulus, $G''(T)$.
The former is a measure of the elastic energy stored per oscillation cycle;
plainly stated, this parameter indicates the degree to which
the sample gives a solid-like response to the dynamic load.
The latter is a measure of the energy dissipated as heat per cycle,
and indicates the degree to which a sample shows liquid-like behaviour.

Empirical measurements were performed using a rheometer (Anton Paar MCR 302).
The samples were thinly sliced and sandwiched between two flat disk plates with a 25 mm radius,
facing each other, separated by a gap of 2 mm.
The sample surface was coated with silicone oil
to prevent evaporation of water during measurements.
After coating the sample, it was gradually cooled,
during which $G'$ and $G''$ were measured by applying the oscillatory shear.
From the $G'$ and $G''$ data, the loss tangent $\tan \delta \equiv G''/G'$
was also evaluated for each temperature and pH condition.

%%%--------------------------------------------------------------
\begin{figure*}[ttt]
\centering
\includegraphics[width=0.45\textwidth]{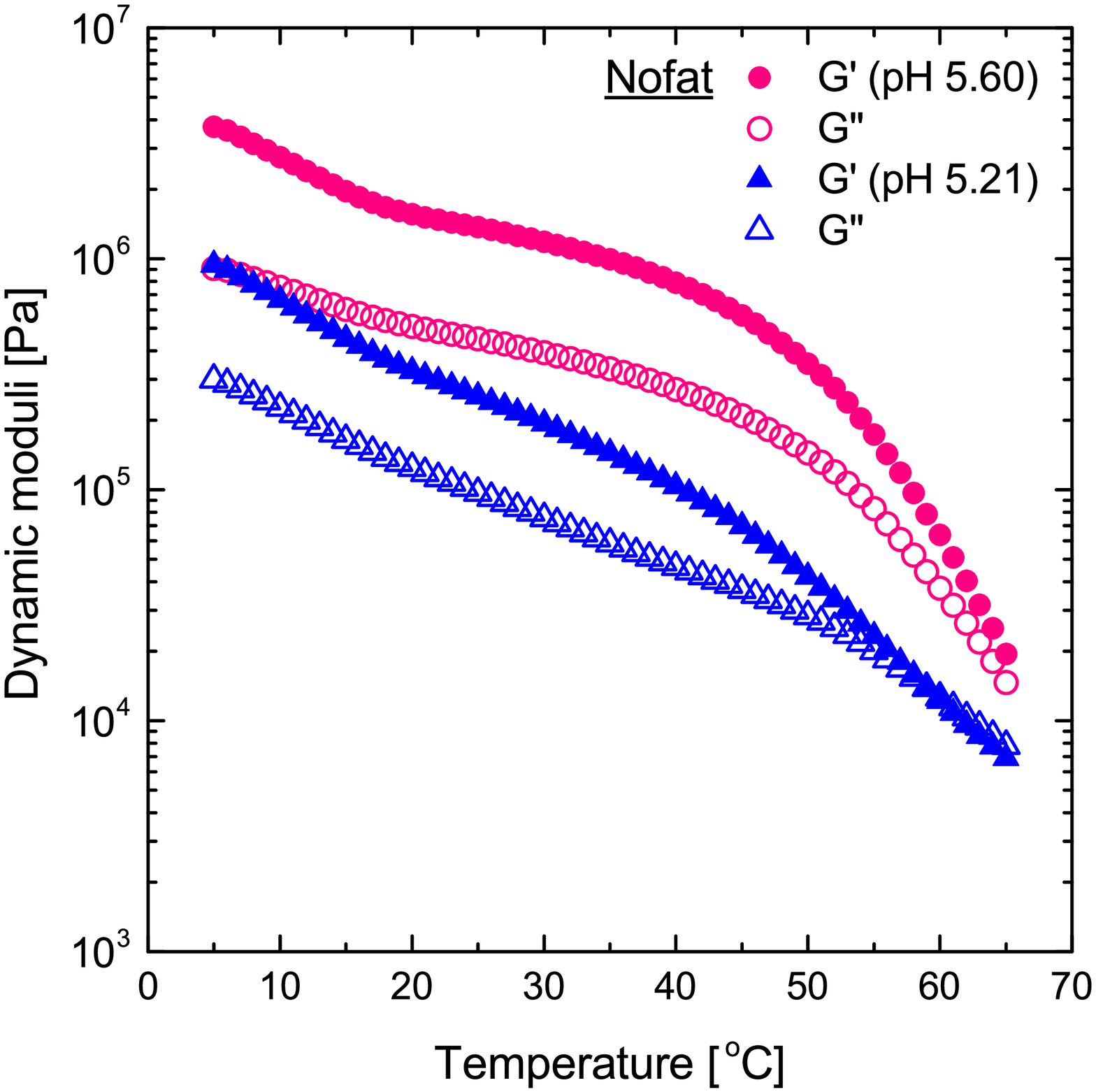}
\hfill
\includegraphics[width=0.45\textwidth]{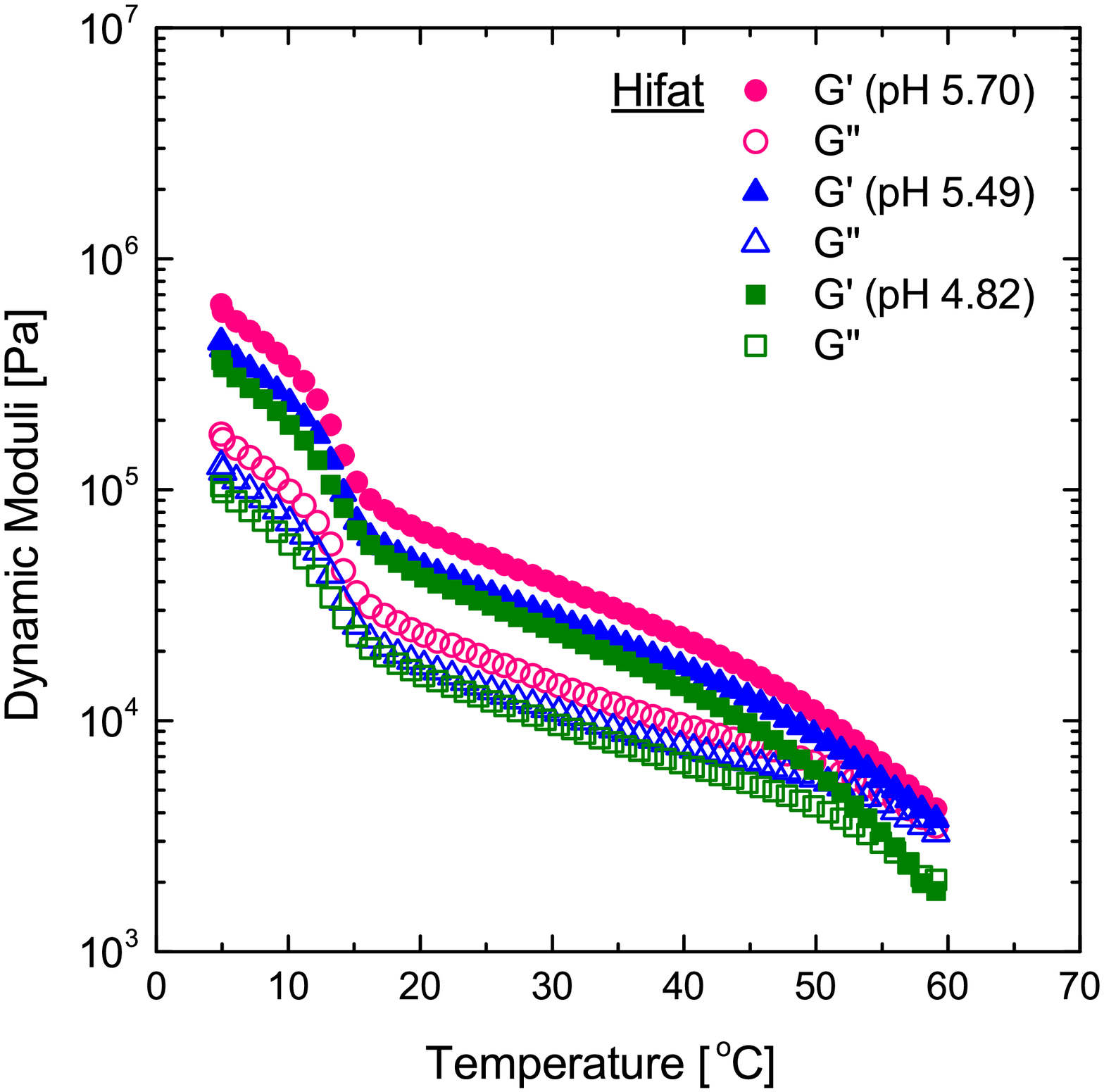}
\caption{Temperature dependence of the dynamic moduli of curds
under different pH conditions.
(a) The no-fat sample; (b) The high-fat sample.}
\label{fig01}
\end{figure*}
%%%--------------------------------------------------------------

\section{Result I: Plasticizer effect and reinforcing effect}

Figure \ref{fig01} shows single-logarithm plots of $G'(T)$ and $G''(T)$
for samples under different pH conditions.
The measured data for the no-fat and high-fat samples are plotted
in Fig.~\ref{fig01}(a) and Fig.~\ref{fig01}(b), respectively.
For each pH condition, 10 samples were analysed and
only minor sample dependence was detected.
Based on the figure, the quasi-static cooling of the samples
from 60 $^\circ$C (or slightly above) to 5 $^\circ$C causes
exponential increases in the magnitude of both $G'(T)$ and $G''(T)$.
This rigidity enhancement driven by slow cooling
is attributed to the disappearance of thermally excited vibrations in the constituents.
By cooling the samples, the degree of thermal vibration in the protein threads as well as
fat globules is depressed
and local detachment between protein threads becomes difficult.
As a result, the samples get firmer as the temperature decreases,
consistent with our observations in our daily life.
This rigidity enhancement seems to be universal for all pH conditions and fat contents.

An effect of fat-control on the magnitudes of $G'(T)$ and $G''(T)$ was clear,
when we compared the high-fat data with the no-fat data presented in Fig.~\ref{fig01}.
At every $T$ and pH,
the moduli for the high-fat samples are nearly one order of magnitude smaller than those for the non-fat samples.
This fat-induced reduction in the moduli is explained by the plasticizer effect of fat globules.
The globules tend to fill in the voids of the protein network
or get between the protein threads.
The insertion of the fat globules into voids or gaps between protein threads
keep them further apart
and reduce the forces of attraction between the threads,
thus making the whole curd more flexible.
Such a plasticizer effect (as well as a possible mechanical cushion effect)
is pronounced at moderately high temperatures ($> 20 ^\circ$C)
because a large portion of the fat in the globules melts and becomes deformable.
Furthermore, the plasticizer effect is enhanced at a relatively large pH ($\sim$ 5.60 - 5.70),
as the protein network is rather sparse and accordingly involves numerous voids into which fat globules can penetrate.

Another important consequence of the fat addition to the raw milk was
a rapid growth in the low-temperature moduli with cooling.
As shown in Fig.~\ref{fig01}(b), below 20 $^\circ$C,
the moduli rapidly increase with cooling for every pH.
This rapid growth in the moduli results from the solidification of fat globules
in the protein matrix.
Below 20 $^\circ$C, the solid domain of the fat pooled in the globules
gradually increases with cooling;
as a result, they begin to function as reinforcing fillers.
A similar reinforcing effect has been observed in our previous study \cite{Shima2015},
in which the viscoelastic moduli of cheese curds free from fat control
were examined.
In the no-fat data shown in Fig.~\ref{fig01}(a), the reinforcing effect disappears almost completely,
and there is only a slight change in the slope of the moduli curves
at approximately 20 $^\circ$C owing to the minimal fat content in the skim milk.

In short, we identified the temperature ranges within which fat globules act as plasticizers and/or reinforcing fillers.
The plasticizing effect was observed
over the whole temperature range (5$^\circ$C -- 60 $^\circ$C),
reducing both the moduli of $G'(T)$ and $G''(T)$ for the high-fat samples.
The reinforcing effect was observed only below 20$^\circ$C,
leading to an excess contribution of the low-temperature moduli
of the high-fat samples.
The identification of these temperature ranges
was the first main result of the present work.

\section{Result II: Sol-gel structural phase transition}

%%%--------------------------------------------------------------
\begin{figure*}[ttt]
\includegraphics[width=0.45\textwidth]{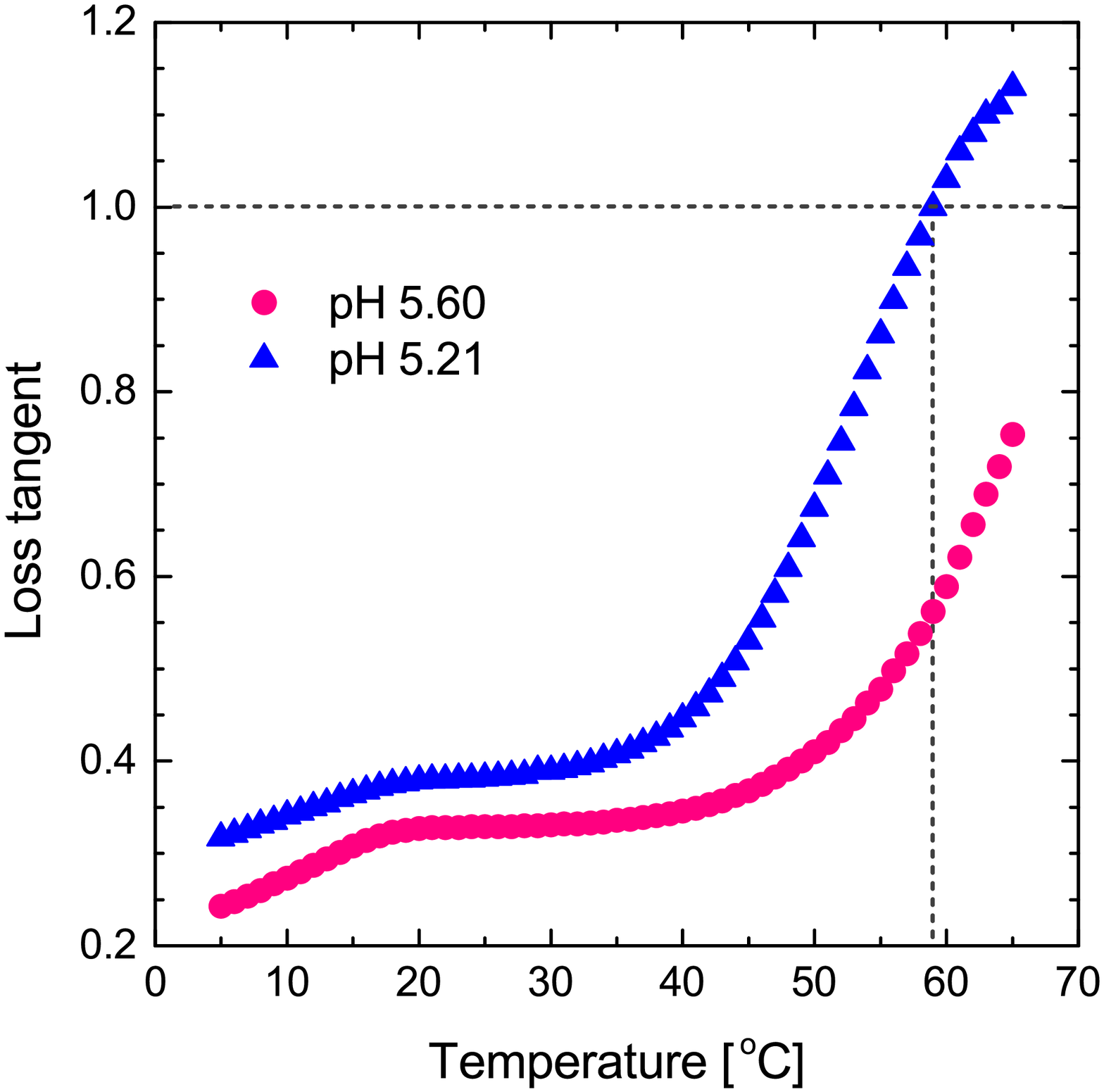}
\hfill
\includegraphics[width=0.45\textwidth]{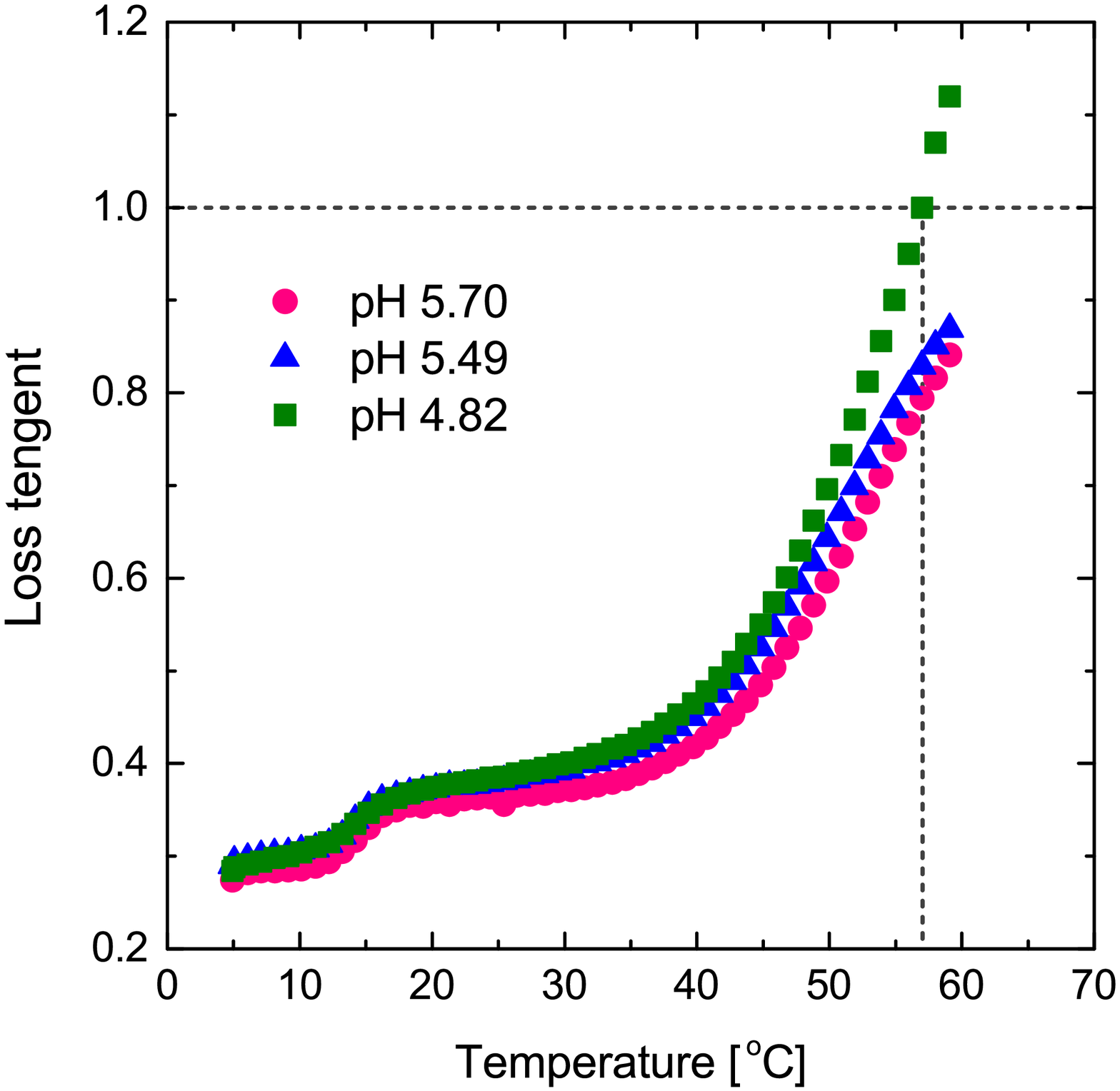}
\caption{Temperature variance of the loss tangent $\tan \delta$ for:
(a) the no-fat case, and (b) the high-fat case.
The sol-gel transition temperature $T_C$ above which $\tan \delta$ exceeds unity
is marked by dotted lines.}
\label{fig02}
\end{figure*}
%%%--------------------------------------------------------------

Figure \ref{fig02} shows the $T$-dependences of the loss tangent, $\tan \delta$,
for the fat-controlled and pH-regulated samples in the same way as those in Fig.~\ref{fig01}.
For both high-fat and no-fat samples, the loss tangent exceeds unity at the following temperature ranges:
$T>59 ^\circ$C for the no-fat case with pH 5.21
and $T>57 ^\circ$C for the high-fat case with pH 4.82.
The high temperature ranges showing $\tan \delta >1$
indicates a structural transition from a sol phase (liquid state)
to a gel phase (solid state) \cite{CYMTung1982,RossMurphy1995},
above which the curds react to an external stress in a more viscous and fluidic,
less elastic manner \cite{Vliet1989}.
The heat-induced flowability is believed to result from higher molecular mobility
and reduced cross-linkage within the casein network \cite{Mleko2005}.
These two (and potentially other) physicochemical factors promote molecular alignment
parallel to the tensile direction, enhancing the flow of the cheese curds at temperatures
that satisfy $\tan \delta > 1$.
The sol-gel transition demonstrated in Fig.~\ref{fig02}
is consistent with the thermal softening of the casein network
in fully coagulated mozzarella cheese reported in the literature \cite{Ak1996}.
Using the squeezing flow method, it has been reported that mozzarella cheese shows
a decreased resistance to flow as the temperature increases; the relaxation time of
mozzarella cheese was reduced by several-fold as the temperature increased
from 30 $^\circ$C to 60 $^\circ$C,
corresponding to a monotonic decrease in $G'(T)$ and $G''(T)$
with increasing $T$ (see Fig.~\ref{fig01}).

It is interesting to examine how the pH and fat content affect the sol-gel phase transition temperature $T_C$.
First, sufficient acidification is required to observe the sol-gel transition.
Reducing the pH promotes the dissociation of calcium ions from the bonding parts of protein molecules.
Hence, the network becomes looser, yielding feasible flow rates above $T_C$.
In our high-fat condition, for instance,
the sample at pH 4.82 undergoes the transition at $T=57 ^\circ$C,
whereas no transition occurs for larger pH values.
In the no-fat condition, a pH of 5.21 suffices for samples to go through the transition at $T=59 ^\circ$C.
Second, the value of $T_C$ decreases as the fat content decreases.
Indeed, our previous study has shown that $T_C=43 ^\circ$C for the sample free from fat-control for a pH of 4.8.
Additionally, based on Fig.~\ref{fig02}(a), for no-fat samples with pH 4.8,
$T_C$ is close to or less than that for fat-control-free samples.
The fat-induced reduction in $T_C$ (for a fixed pH) indicates
the consistent adhesion property of fat globules to the protein matrix.
Since liquefied globules that are soft and deformable tend to maintain adhesion
to surrounding protein molecules,
adjacent protein threads are glued and cannot break apart.
If the fat content decreases, the gluing mechanism is suppressed and the gel phase is favoured
at moderately high temperatures.
This explains why $T_C$ values for the natural-fat and no-fat samples are smaller than that of the high-fat samples.

In short, we revealed the effect of fat content on the sol-gel transition temperature $T_C$.
The addition of excess fat to raw milk results in an upward shift
of the $T_C$ of cheese curds due to the adhesion property of fat globules
to protein threads.
This is the second main result of the present article.

\section{Conclusion}

We investigated the effect of fat content variation
on dependences of $G'(T)$ and $G''(T)$
on $T$ for mozzarella-type cheese curds.
We observed two distinct effects.
Specifically, we detected a fat-induced reduction in the moduli ({\it i.e.,} the plasticizing effect)
at all temperatures between 5$^\circ$C  and 60 $^\circ$C,
and a fat-induced excess contribution to the moduli
({\it i.e.,} the reinforcing effect)
that is observable only below 20$^\circ$C.
The former effect is attributed to the presence of liquefied fat globules
wrapped around the three-dimensional protein network,
which reduce the attractive forces between the threads,
making the system more flexible.
With additional cooling, in contrast, a portion of fat pooled in the globules
becomes solidified and begins to function as a reinforcing filler;
this results in the latter effect.
In addition to these two effects,
we revealed the fat-induced reduction in
the critical temperature $T_C$ for the sol-gel phase transition
of the cheese curds.
This increase in $T_C$ indicates the adhesion property of
liquefied fat globules to protein threads.

\begin{acknowledgements}
The authors express their gratitude to Emeritus Prof.~Ryoya Niki, Prof.~Katsuyoshi Nishinari,
Prof.~Kaoru Sato, and Mr.~Kunio Ueda
for fruitful discussions and technical supports.
This work was supported by JSPS KAKENHI Grant Numbers 25390147 and 25560035.
\end{acknowledgements}

\noindent
{\small
{\bf Conflict of interest} $\quad $None.
}

\vspace*{8pt}

\noindent
{\small
{\bf Compliance with Ethics Requirements} $\quad$
This article does not contain any studies with human or animal subjects.
}

\bibliographystyle{spphys}       % APS-like style for physics

\bibliography{HShima_fatcontrol_EFRT.bib}   % name your BibTeX data base

%************************
\end{document}